# Hump Structure below $T_c$ in the thermal conductivity of $MgB_2$ superconductor

R. Lal, Arpita Vajpayee, V. P. S. Awana, and H. Kishan
National Physical Laboratory, Dr. K.S. Krishnan Marg, New Delhi-110012, India

A. M. Awasthi
UGC-DAE Consortium for Scientific Research, University Campus, Khandwa Road, Indore-452017, India

## ABSTRACT

A reasonable cause of absence of hump structure in thermal conductivity of $MgB_2$ below the superconducting transition temperature ($T_c$) lies in the appearance of multigap structure. The gaps of lower magnitude can be suppressed by defects so that this system becomes effectively a single gap superconductor. When such a situation is created, it is hoped that thermal conductivity ($\kappa$) will show hump below $T_c$. Proceeding along these lines, a sample of $MgB_2$ with a relatively higher residual resistivity $\rho_o$ = 33.8 $\mu\Omega$-cm has been found to show a hump structure below $T_c$. The actual electronic thermal conductivity $\kappa_{el}$ of this sample is less than that expected from the Wiedeman- Franz law by more than a factor of 2.6 in the considered temperature range. Modifying the Wiedeman- Franz law for the electronic contribution by replacing the Lorenz number $L_0$=2.45x10$^{-8}$ W$\Omega$K$^{-2}$ by an effective Lorenz number $L_{eff}$ (<$L_0$) we have obtained two sets of $\kappa_{el}$, namely those with $L_{eff}$ = 0.1$L_0$ and 0.2$L_0$. Corresponding to these two sets of $\kappa_{el}$, two sets of the phonon thermal conductivity $\kappa_{ph}$ are obtained. $\kappa_{ph}$ has been analyzed in terms of an extended Bardeen- Rickayzen- Tewordt theory. The main result of this analysis is that the hump structure corresponds to a gap ratio of 3.5, and that large electron-point defect scattering is the main source of drastic reduction of the electronic thermal conductivity from that given by the usual Wiedeman- Franz law.

# I. INTRODUCTION

The $MgB_2$ superconductor was discovered by Nagamatsu et al [1] in 2001. It involves two types of carriers – one corresponding to the two dimensional (2D) $\sigma$ band, and the other corresponding to the three-dimensional (3D) $\pi$ band [2]. The dominant effect of superconductivity takes place only in the $\sigma$ band [2], although $\pi$ band also takes part in the superconductivity [3]. The interaction responsible for the superconductivity is the electron-phonon interaction [2, 3]. In the normal state the resistivity of the $MgB_2$ metal is also caused by the electron-phonon interaction in a defect-free sample [4]. There has been an intensive study of almost all the types of physical properties of the $MgB_2$ superconductor in both the super conducting and normal states, and thus sufficient progress has been made to understand the behavior of this compound [5]. In particular, the thermal conductivity has been investigated by a large number of workers [6-16] by considering various forms of the $MgB_2$ system, like different defect levels, different types of doping, and single or polycrystalline samples. The studies of the thermal conductivity ($\kappa$) of $MgB_2$ made by Sologubenko et al [9,13], Wu et al [15] and Anshukeva et al [16] is limited to low temperatures ($T \leq 100$K) only and in these studies $\kappa$ increases with temperature $T$ within the considered temperature range. Other authors [6-8, 10-12, 14] have studied the thermal conductivity up to much higher temperature range ($T \leq 250$K [7], $T \leq 275$K [10, 11], $T \leq 300$K [6, 8, 12, 14]). The general behavior of $\kappa$ with $T$ is that initially $\kappa$ increases with $T$ attaining a maximum value at some temperature $T_{max}$. The values of $T_{max}$ are 112 K, 118K, 66K, 66K and 66K for the samples of References [6], [7], [8], [10] and [11] respectively. The behavior of the $MgB_2$ sample of Ref. [12] is quite complicated, and we have considered only the MgB11-15 sample of Ref. [11]. The samples of Ref. [14] does not show any maximum in $\kappa$ till $T$ = 300K. The MGB-TS sample of Ref. [10] also does not show any maximum till 275K. Another general feature of $\kappa$ is that in some cases [6,7,14], $\kappa$ increases with $T$ near room temperature, while in another cases [8,11] κ continues to decrease after $T_{max}$ up to the highest measured temperature.

It is well established that the superconducting state of the $MgB_2$ superconductor shows two finite gaps at the Fermi energy at $T$ = 0K [17, 20]. One of these gaps corresponds to the σ band, while the other to the π band. From a superconductor of finite gap(s) we expect according to BRT theory [21, 22] a hump below $T_c$ in the thermal conductivity. But a weak hump is reported only by Anshukeva et al [16] below $T_c$, as none of the other existing reports [6-15] show a hump structure in thermal conductivity below $T_c$. Various authors interperate the absence of a hump structure in κ below $T_c$ in various ways. It has been argued in particular by Sologubenko et al [8] that absence of the hump structure is possible if the energy gap is about three times smaller than the values given by the original Bardeen-Cooper-Schriffer (BCS) theory [23]. But the reported energy gap at zero temperature, 2 Δ(0) is found to be 4.1 $k_B T_c$ by Chen et al [17], and 4.3 $k_B T_c$ by Lui et al [3] for the σ band. This means that the superconducting energy gap is certainly much larger than one third of the BCS gap $2\Delta_{BCS}(0) = 3.5 k_B T_c$. Other possible reason for absence of the hump structure is that the electron-phonon interaction is much weaker than phonon-defect scattering. This is shown to be incompatible with the situation that exists in $MgB_2$ [8]. Yet another possible reason is that the phonon contribution to the thermal conductivity is small near $T_c$. This is also shown incompatible with the situation of $MgB_2$ [8]. At present the most reasonable source of absence of hump structure below $T_c$ in $MgB_2$ is the possibility $MgB_2$ is a multigap superconductor [3, 8].

The multigap structure of the superconductivity of $MgB_2$ may correspond to the clean limit [3] or dirty limit [24] depending upon the nature of the sample. For a $MgB_2$ sample of residual resistivity $\rho_o$ = 2.0 μΩ-cm, Sologubenko et al [8] find $\ell \approx$ 800Å. The coherence length of the polycrystalline sample of $MgB_2$ is $\xi \approx$ 52 Å [25]. This means $MgB_2$ will no more be in the clean limit if $\rho_o$ is enhanced by a factor of 15 or more, i.e. if $\rho_o \geq$ 30μ Ω-cm. When this condition is met we hope the multigap nature of the superconducting state to change to a single-gap case, thereby making it possible to observe a hump in κ vs $T$ below $T_c$. On these lines we have prepared a sample of $MgB_2$ with $\rho_o$ = 33.8 μ Ω-cm, and have indeed observed a clear hump in κ vs $T$ curve below $T_c$. Here it may however be noted that the MGB-TS sample of Putti et al [10] corresponds to $\rho_o$ = 39.0 μ Ω-cm. Despite this these authors have not observed any hump in the thermal conductivity below $T_c$.

## II. EXPERIMENTAL DATA

Synthesis of polycrystalline bulk sample of $MgB_2$ is described by Awana et al [26]. Micro-structural details and measurements of various physical properties, including X-Ray diffraction, magnetization, electrical resistivity and thermal conductivity are also described in Ref. [26]. For clarity here we reproduce the behavior of XRD, magnetization, electrical resistivity and thermal conductivity in Figs. 1 and 2. The resistivity has been measured up to 300K and the thermal conductivity has been measured from 18K to 300K. The micro-structural analysis of Awana et al [26] shows that the grains of $MgB_2$ have excellent connectivity, and in particular there is no evidence of cracks in the system. Thus the present thermal conductivity data corresponds to the actual behavior of the $MgB_2$ sample. The fact that the observed thermal conductivity appears to be lowest than those of other authors [6–16] does not point to any inconsistency. This is because the residual resistivity $\rho_o$ is much higher in the present case than most of the other reports [6-16]. In fact, we found that the present value of $\rho_o$ is comparable to the MGB-TS sample of Putti et al [10] where $\rho_o$ = 39.0 μ Ω-cm. Thus we should compare our κ results with those of these authors only. When we do so, it turns out that the values of κ are 3.1 W/m K and 4.5 W/m K at $T$ = 25K and 100K respectively in the present case, while the corresponding values of κ for the MGB-TS sample of Putti et al [10] are 2.0 W/m K and 6.1 W/m K. Obviously these values are comparable, signifying that low value of κ in the present case are due to higher $\rho_o$, and not due to cracks etc.

## III. RESULTS AND DISCUSSION

From the magnetization of Fig. 1 we find that the present sample of $MgB_2$ superconducts at $T_c$ = 37.3K. From the peak of the slope $d\rho/dT$ of $\rho$ vs $T$ curve of Fig. 2 we find that $T_c$ = 37.4K. The hump structure in the thermal conductivity also corresponds to $T_c$ = 37.4K. In the following description we will for specificity take $T_c$ = 37.4K. The resistivity of Fig. 1 corresponds to a residual resistivity of $\rho_o$ = 33.8 μΩ-cm. The values of the residual resistivity ($\rho_o$) for the samples, considered by the authors of Refs. [11], [8], [6] and [7] are 0.56, 2.0, 12.5 and 12.5μ Ω-cm respectively, and the corresponding values of $T_c$ are 38.6, 38.1, 37.5 and 37.5K. When we compare these $\rho_o$ vs $T_c$ values of various authors [6-8, 11], along with the present values, it turns out that the lower value of $T_c$ in the present case is due to higher value of $\rho_o$. If superconductivity is caused by the electron-phonon interaction, sample disorder will enhance the normal self-energy, which in turn, reduces $T_c$ [27].

Following the two band analysis of the resistivity of Ref. [14] in the present case we find that the considered $MgB_2$ sample corresponds to the Debye temperature $\theta_D$ = 1024K, and to the residual values of the resistivity for σ and π bands, $\rho_{0,\sigma}$ = 60.7 μΩ-cm and $\rho_{0,\pi}$ = 76.2 μ Ω-cm, respectively.

We now turn to the thermal conductivity κ. It is a sum of the electronic contribution $\kappa_{el}$ and the phonon contribution $\kappa_{ph}$. That is to say

$$\kappa = \kappa_{el} + \kappa_{ph} \tag{1}$$

First we consider the electronic contribution $\kappa_{el}$. According to Wiedemann- Franz law, for elastic scattering processes, the thermal conductivity is given by

$$\kappa_{el} = L_0 T / \rho(T) \tag{2}$$

Here $L_0$ = 2.45x $10^{-8}$ $W\Omega K^{-2}$ is the Lorenz number, and $\rho(T)$ is the resistivity of the $MgB_2$ sample. Eq. (2) has been used by many workers [6, 8-11, 13, 14] for the estimation of the electronic thermal conductivity. The values of $\kappa_{el}$, calculated on the basis of Eq. (2) are shown in Fig. 2. It is seen from this figure that $\kappa_{el} < \kappa$ for only $T <$ 52K. As the temperature increases beyond 52K, $\kappa_{el}$ becomes higher and higher than $\kappa$ such that $\kappa_{el}$ = 9.29 W/m K as compared to $\kappa$ = 3.58 W/m K at 300K. Since according to Eq. (1) $\kappa_{el}$ should not exceed $\kappa$ at any temperature, Eq. 2 is inadequate for the description of the electronic thermal conductivity. The actual electronic thermal conductivity should be considerably smaller e.g. by a factor of more than 2.6 at 300K. This means that the Wiedemann- Franz law is drastically affected in the present case. On the other hand this law is well applicable for the $MgB_2$ samples MB1 and MB2 considered by Sologubenko et al [9]. The main difference between the present $MgB_2$ sample and those of Sologubenko et al lies in the much different value of $\rho_o$. While the MB1 and MB2 samples of Sologubenko correspond to $\rho_o$< 0.71 and $\rho_o$< 1.20 μΩ-cm respectively, the present sample correspond to $\rho_o$ = 33.8 μΩ-cm.

One of the possible effect of higher $\rho_o$ is that the system involves localization process. In order to see how far this is possible we note that Sologubenko et al [8] have found that the mean free path for $\rho_o$ = 2.0 μΩ-cm is about 800Å. In the present case $\rho_o$ = 33.8 μΩ-cm. So we expect $\ell \approx 50$ Å. Since, the Fermi velocity $v_F$ of $MgB_2$ is $4.9\times10^7$cm/sec [8], we obtain $k_F\ell \approx 20$. This value is large enough to keep the system much away from the localization [28]. Another possible effect of large $\rho_o$ is a movement of the mobility edge [28] towards the Fermi level. This will effectively reduce the Fermi energy so that the parameter $(k_BT/E_F)^2$ no more satisfies the condition $(k_BT/E_F)^2 \ll 1$. In fact, $(k_BT/E_F)^2$ is the expansion parameter for obtaining the Lorenz number. On the basis of pages 217 and 676 of Mahan [29], it may be shown that the first-order correction in the expansion parameter $(k_BT/E_F)^2$ is negative with a magnitude $(\kappa_{el}/3)(k_BT/E_F)^2$. This means that the first-order correction will reduce the Lorenz number. The Fermi energy corresponding to the σ band of $\rho_o \approx 0$ μΩ-cm $MgB_2$ is about 0.7 eV or about 8000K [2]. Thus in $MgB_2$ the condition for the first-order correction to be negligible demands $T^2 \ll$ 8000 or $T \ll$90K. This means that even for $\rho_o \approx 0$ μΩ-cm $MgB_2$ the Wiedemann-Franz law will be modified for $T \geq 90$K. Fig. 4 of Putti et al [11] shows that Wiedemann-Franz law is modified for $\rho_o < 2.1$ μΩ-cm sample even at 50K. Sologubenko et al [9] have found that the Wiedemann-Franz law is well satisfied in various $MgB_2$ samples corresponding to $\rho_o < 4.17$ μΩ-cm for $T \leq 6$K. This temperature range and $\rho_o$ are too low to violate the condition $(k_B T/E_F)^2 \ll 1$. Also the phonon contribution to thermal conductivity for $T \leq 6$K is too low since $\kappa_{ph} \sim T^3$ (for low $T$). Combining all these factors we argue that for $T \leq 6$K, the first-order reduction to $\kappa_{el}$ and the phonon thermal conductivity $\kappa_{ph}$ are negligible so that validity of the Wiedemann-Franz law is well expected. This is why Sologubenko et al [9] found the Wiedemann-Franz law to be valid. When the Fermi energy reduces due to the mobility edge effect in the presence of higher disorder in a system [28], the validity of the Wiedemann-Franz law will certainly be affected. Thus a reduction of $\kappa_{el}$ by a fraction of 10 or so is possible in $MgB_2$ sample involving large amount of (point) defects.

In view of the above we describe the electronic contribution to the thermal conductivity by

$$\kappa_{el} = L_{eff}\, T / \rho(T) \tag{3}$$

such that $L_{eff} < L_0$. We emphasize that the effective Lorenz number $L_{eff}$ corresponds to electrons and does not involve the contribution of the phonons. This specification of $L_{eff}$ is different from those considered earlier in literature by other authors [9, 10]. Thus in the present case we never expect $L_{eff}$ to be larger than $L_0$.

From the above-mentioned comparison of $\kappa_{el}$ and $\kappa$ at 300K it may be said that $L_{eff}$ should be essentially less than $0.38L_0$. For specificity we take two values of $L_{eff}$, $0.1L_0$ and $0.2L_0$, and then obtain the phonon contribution to thermal conductivity $\kappa_{ph}$ from Eq. (1) for $T > T_c$. For $T < T_c$ we use Fig.1 of BRT [21] for estimating values of $\kappa_{el}$. The values of $\kappa_{ph}$ obtained in this manner are shown in Fig. 3.

In order to analyze the phonon thermal conductivity $\kappa_{ph}$ we employ the following expression [22, 30], which is based on the BRT theory [21]

$$\kappa_{ph} = t^3 \int_0^{\theta_D / T} \frac{x^4 e^x}{(e^x - 1)^2 S(t,x)} dx \quad (4)$$

Here $t = T/ T_c$ is reduced temperature and the scattering rate $S(t,x)$ is given by

$$S(t, x) = S_1 + S_2 t^4 x^4 + S_3 t^2 x^2 + S_4 tx + S_5 tx \; g(x,\Delta(t)/ k_B t T_c) + S_6 t^4 x^4 \quad (5)$$

Here $S_1$ is, to within a constant (say C), phonon-boundary scattering rate. $S_2$, $S_3$, $S_4$, $S_5$ and $S_6$ are to within C the scattering rates of phonon with point defects, sheet like faults, dislocation, electrons and phonon respectively [22, 30]. The function $g(x,\Delta(t)/k_B t T_c$ is defined extensively by Tewordt and Wolkhausen [22].

Taking the value of the Debye temperature $\theta_D$ from the analysis of the resistivity ($\theta_D$ = 1024K) we have fitted the two sets of $\kappa_{ph}$ values of Fig. 3 by Eq. (4). Various parameters are presented in table-I. We see that the strain field of sheet-like faults provides highest scattering rate, while the strain field of dislocations corresponds to lowest scattering rate. The superconducting gap ratio $2\Delta(0)/k_B T_c$ is found to be 3.5, near the BCS value [23] for both the values of $L_{eff}$. The value of $2\Delta(0)/k_B T_c$ reported in literature is as high as 4.3[3]. The large value of disorder in the present case might have reduced the gap ratio from 4.3 to 3.5.

From table-I we find that the contribution of point defects ($S_2$) varies by about 22% for the variation of $L_{eff}$ from $0.1L_0$ to $0.2L_0$. The corresponding variation for the phonon-phonon scattering ($S_6$) is about 10%. The variations of $S_1$, $S_3$, $S_4$ and $S_5$ are negligibly small. The phonon-phonon scattering is limited to the behavior of phonon only. So we cannot connect it with the electronic carriers. On the other hand, the point defects will scatter the electronic carriers also. Thus we may say that the drastic suppression of the values of $\kappa_{el}$ in the present case from that of the $\kappa_{el}$ values given by Wiedemann-Franz law (Eq.2) is due to point defects. Since the non-appearance of hump structure in $\kappa$ below $T_c$ is believed to be due to multigap nature of the superconductivity, we may argue that the intensive effect of the point defects has suppressed the $\pi$-gap so that there remains only one gap in the present system. When this is so, the reason of the non-appearance of a hump below $T_c$ in the thermal conductivity of the MGB-TS sample of Putti et al [10] is that this sample has not yet been reduced to a single gap superconductor.

## IV. CONCLUSIONS

Before planning the study of this paper we have investigated various possible reasons for the missing of a hump structure in the thermal conducting below $T_c$. The main guiding point for $MgB_2$ samples showing hump structure below $T_c$ in the $\kappa$ vs $T$ behavior was believed to convert the multigap structure of the superconducting state to an effectively single-gap structure. The trick for performing this task was realized to introduce sufficient defects in the system. Since the hump structure appears due to finite superconducting gap, and since this gap may also be suppressed by large defects, we prefer to prepare a $MgB_2$ system with such defect levels which can provide $\rho o \approx 30$ μΩ-cm. The $MgB_2$ sample prepared by us shows a clear hump structure in $\kappa$ of this sample below $T_c$.

We have found that the electronic contribution to thermal conductivity, as calculated on the basis of the Wiedemann-Franz law, exceeds the observed thermal conducting for $T \geq 52$ K. This means that the present sample corresponds to a reduced Lorenz number ($L_{eff} < 2.45\text{x}10^{-8}$ W Ω $K^{-2}$). Some authors (e.g. Ref. [11]) have found such a situation earlier also. In order to analyze the thermal conductivity, we have taken two values of the effective Lorenz number, $L_{eff} = 0.1L_0$ and $0.2L_0$, for estimating the electronic thermal conductivity. The reason of such low values of $\kappa_{el}$ has been argued in terms of reduced Fermi energy due to mobility edge effect of the defects. In fact, when $E_F$ reduces the validity of the Wiedemann-Franz law shifts towards much lower value of $T$.

Having specified the electronic thermal conductivity in a phenomenological manner, we have estimated the lattice thermal conductivity $\kappa_{ph}$ by using Eq. (1). The two sets of the resulting thermal conductivity have been analyzed in terms of a theory based essentially on the Bardeen- Rickayzen –Tewordt formulation. The analysis provides two main results. First the gap ratio of the present sample turns out to be 3.5, which is lower than that corresponding to the σ band in the clean limit. A possible reason for this is that defects have reduced the gap ratio. Another outcome of the analysis of the thermal conductivity data is that the drastic reduction of the electronic thermal conductivity is driven by the point defects.

## V. ACKNOWLEDGEMENT

The authors from *NPL* would like to thank Dr. Vikram Kumar (Director) for his continuous encouragement in present work.

## REFERENCES

1. J. Nagamatsu, N. Nakagawa, T. Muranaka, Y. Zenjtani and J. Akimitsu, Nature (London), 410, 63(2001)

2. J. M. An and W. E. Pickett, Phys. Rev. Lett. 86, 4366 (2001)

3. Amy Y. Liu, I. I. Mazin and J. Kortus, Phys. Rev. Lett. 87, 087005 (2001)

4. I. I. Mazin, O. K. Andersen, O. Jepsen, O. V. Dolgov, J. Kortus, A. A. Golubov, A. B. Kuz'menko and D. van der Marel, Phys. Rev. Lett. 89, 107002 (2002)

5. J. Kortus, Physica C 456, 54 (2007)

6. E. Bauer, Ch. Paul, St. Berger, S. Majumdar, H. Michor, M. Giovannini, A. Saccone, and A. Bianconi, Journal of Physics: Cond. Matt. 13, L487 (2001)

7. T. Muranaka, J. Akimitsu and M. Sera, Phys. Rev. B 64, 020505 (2001)

8. A. V. Solugobenko, J. Jun, S. M. Kazakov, J. Karpinski and H. R. Ott, Phys. Rev. B 66, 014504 (2002)

9. A. V. Solugobenko, N. D. Zhigadlo, J. Karpinski, and H. R. Ott, Phys. Rev. B. 74, 184523 (2006)

10. M. Putti, V. Braccini, E. Galleani, F. Napoli, I. Pallecchi, A. S. Siri, P. Manfrinetti and A. Palenzona, Supercond. Sci. Technol. 16, 188 (2003)

11. M. Putti, V. Braccini, E. Galleani d'Agliano, F. Napoli, I. Pallecchi, A. S. Siri,

P. Manfrinetti and A. Palenzona, Phys. Rev. B 67, 064505 (2003)

12. J. Mucha, M. Pekala, J. Szydlowska, W. Gadomski, J. Akimitsu, J-F Fagnard, P. Vanderbemden, R. Cloots and M. Ausloos, Supercond. Science and Tech. 16, 1167 (2003)

13. A. V. Solugobenko, N. D. Zhigadlo, S. M. Kazakov, J. Karpinski and H. R. Ott, Phys. Rev. B 71, 020501 (2005)

14. B. Gahtori, R. Lal, S. K. Agarwal, Y. K. Kuo, K. M. Sivakumar, J. K. Hsu, J. Y. Lin, A. Rao, S. K. Chen and J. L. MacManus-Driscoll, Phys. Rev. B 75, 184513 (2007)

15. Bai- Mei Wu, Dong-Sheng Yang, Wei-Hua Zeng, Shi-Yan Li, Bo Li, Rong Fan, Xian-Hui Chen, Lie-Zhao Cao and Marcel Ausloos, Supercond. Science and Tech. 17, 1458 (2004)

16. N. V. Anshukova, B. Bulychev, A. Golovashkin, L. Ivanova, A. Minakov, A. Rusakov, Physics of the solid state 45, 1207 (2003)

17. X. K. Chen, M. J. Konstaninovic, J. C. Irwin, D. D. Lawrie and J. P. Franck, cond-mat 0104038 (2001)

18. A. Plecenik, S. Benacka and P. Kus 2001, cond-mat/0104038

19. R.S. Gonnelli, D. Daghero, G.A. Ummarino, M. Tortello, D. Delaude, V.A. Stepanov, J. Karpinski, Physica C 456, 134 (2007)

20. V. Ferrando, M. Affronte, D. Daghero, R. Di Capua, C. Tarantini, M. Putti, Physica C 456, 144 (2007)

21. J. Bardeen G. Rickayzen and L. Tewordt, Phys. Rev. 113, 982 (1959)

22. L. Tewordt and Th. Wolkhausen, Solid State Commu. 70, 839 (1989)

23. J. Bardeen, L. N. Cooper andJ. R. Schrieffer, Phys. Rev. 108, 1175 (1957)

24. V. Braccini, A. Gurevich, J. E. Giencke, M. C. Jewell, C. B. Eom, D. C. Larbalestier, A. Pogrebnyakov, Y. Cui, B. T. Liu, Y. F. Hu, J. M. Redwing, Qi Li, X. X. Xi, R. K. Singh, R. Gandikota, J. Kim, B. Wilkens, N. Newman, J. Rowell, B. Moeckly, V. Ferrando, C. Tarantini, D. Marré, M. Putti, C. Ferdeghini, R. Vaglio, and E. Haanappel, Phys. Rev. B 71, 012504 (2005)

25. D. K. Finnemore, J. E. Ostenson, S. L. Bud'ko, G. Lapertot and P. C. Canfield, Phys. Rev. Lett. 86, 2420 (2001)

26. V. P. S. Awana, Arpita Vajpayee, Monika Mudgel, V. Ganesan, A.M. Awasthi, G.L. Bhalla, and H. Kishan, Eur. Phys. J. B 62, 281 (2008)

27. D. Belitz, Phys. Rev. B 36, 47 (1987)

28. P. A. Lee and T. V. Ramakrishnan, Rev. Mod. Phys. 57, 287 (1985)

29. G. D. Mahan, Many-particle Physics, I$^{st}$ Edition (Plenum, New York, 1981)

30. S. D. Peacor, R. A. Richardson, F. Nori and C. Uher, Phys. Rev. B 44, 9508 (1991)

Table I.

Values of various scattering rates $S_1$, $S_2$, $S_3$, $S_4$, $S_5$ and $S_6$, and the gap ratio= $2\Delta(0)/\ k_B T_c$ for the effective Lorenz numbers, $L_{eff} = 0.1L_0$ and $0.2L_0$. To Within a constant (C), the various scattering rates of phonons are with boundary ($S_1$), point defects ($S_2$), strain field of sheet like faults ($S_3$), strain field of dislocations ($S_4$), electron ($S_5$) and phonons ($S_6$). The units of $S_i$ are m K/kW for all i = 1,2,…..6.

| $L_{eff}$ | $S_1$ | $S_2$ | $S_3$ | $S_4$ | $S_5$ | $S_6$ | $2\Delta(0)/\ k_B T_c$ |
|---|---|---|---|---|---|---|---|
| $0.1L_0$ | 31.3 | 172 | 250 | 21.8 | 56.2 | 24.4 | 3.5 |
| $0.2L_0$ | 31.5 | 221 | 252 | 22.1 | 56.7 | 26.8 | 3.5 |

## FIGURE CAPTIONS

Fig. 1: Observed resistivity $\rho$(T) of the $MgB_2$ sample up to 300K. The upper inset shows the X- ray diffraction, and the lower inset shows the magnetization $M$ for $T \leq 45$K.

Fig. 2: Observed thermal conductivity $\kappa$ of the $MgB_2$ sample from 18K to 300K. The dashed line corresponds to the electronic thermal conductivity $\kappa_{el} = L_o T/\rho(T)$. The inset shows the thermal conductivity near the superconducting transition temperature $T_c$.

Fig. 3: Values of phonon thermal conductivity $\kappa_{ph}$ for $L_{eff} = 0.1L_0$ and $0.2L_0$. The values of $\kappa_{ph}$ for $T < T_c$ are obtained by taking $\kappa_{el}$ from Fig.1 of Ref. 17. The symbols corresponds to the experimental values, while the solid lines correspond to the values of $\kappa_{ph}$ calculated according to Eq.(4).

Fig. 1 R. Lal et al

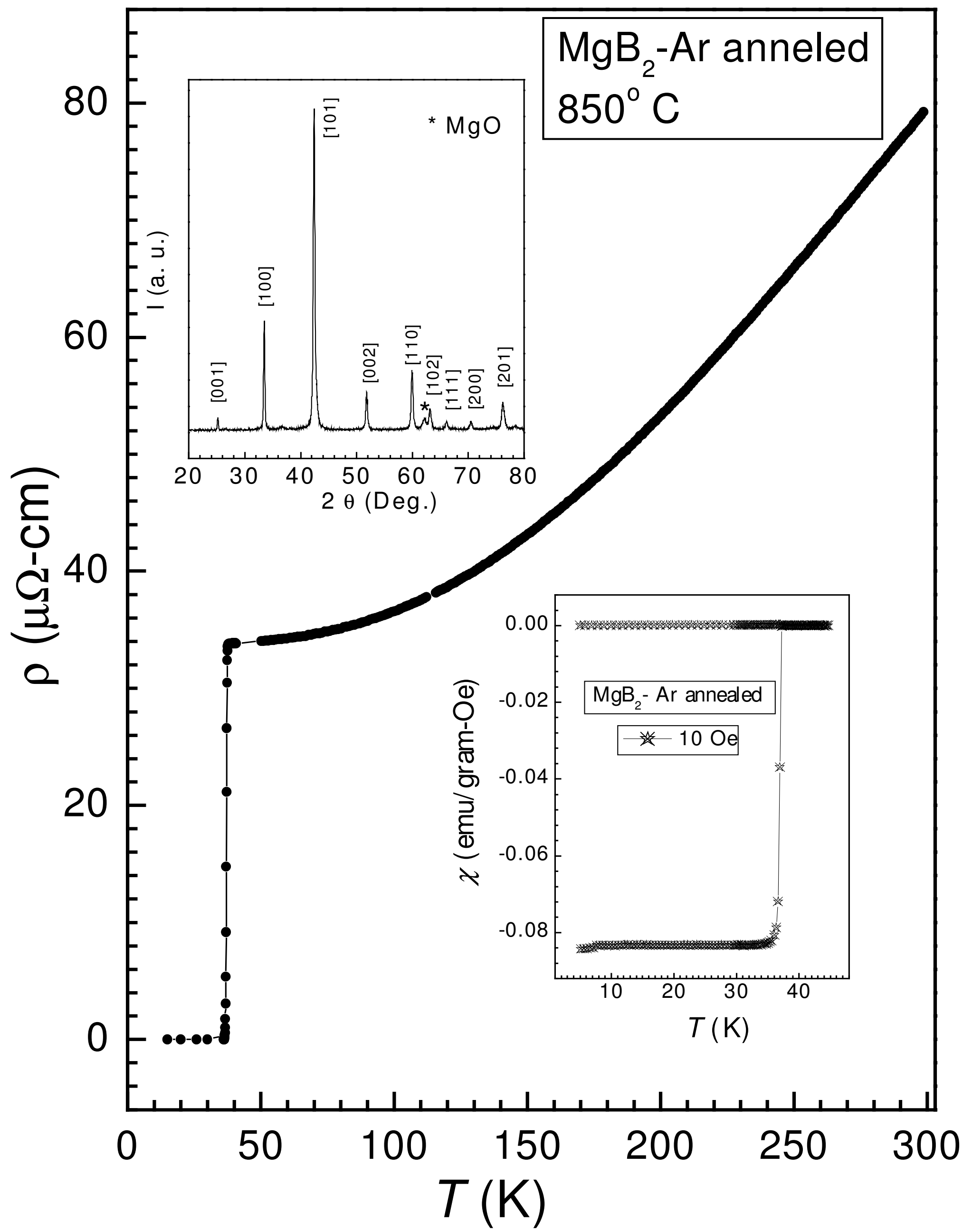

Fig. 2 R. Lal et al

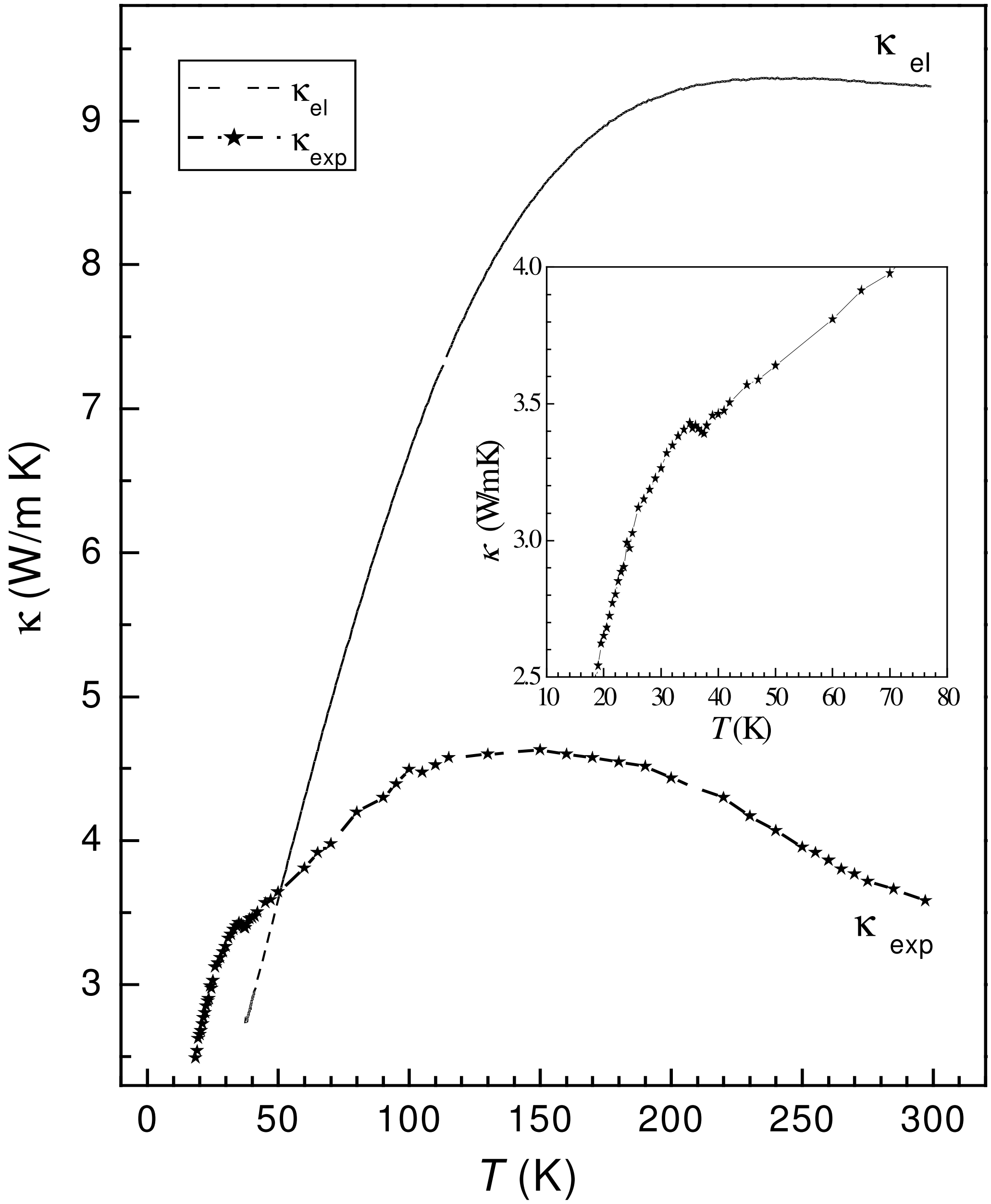

Fig. 3 R. Lal et al

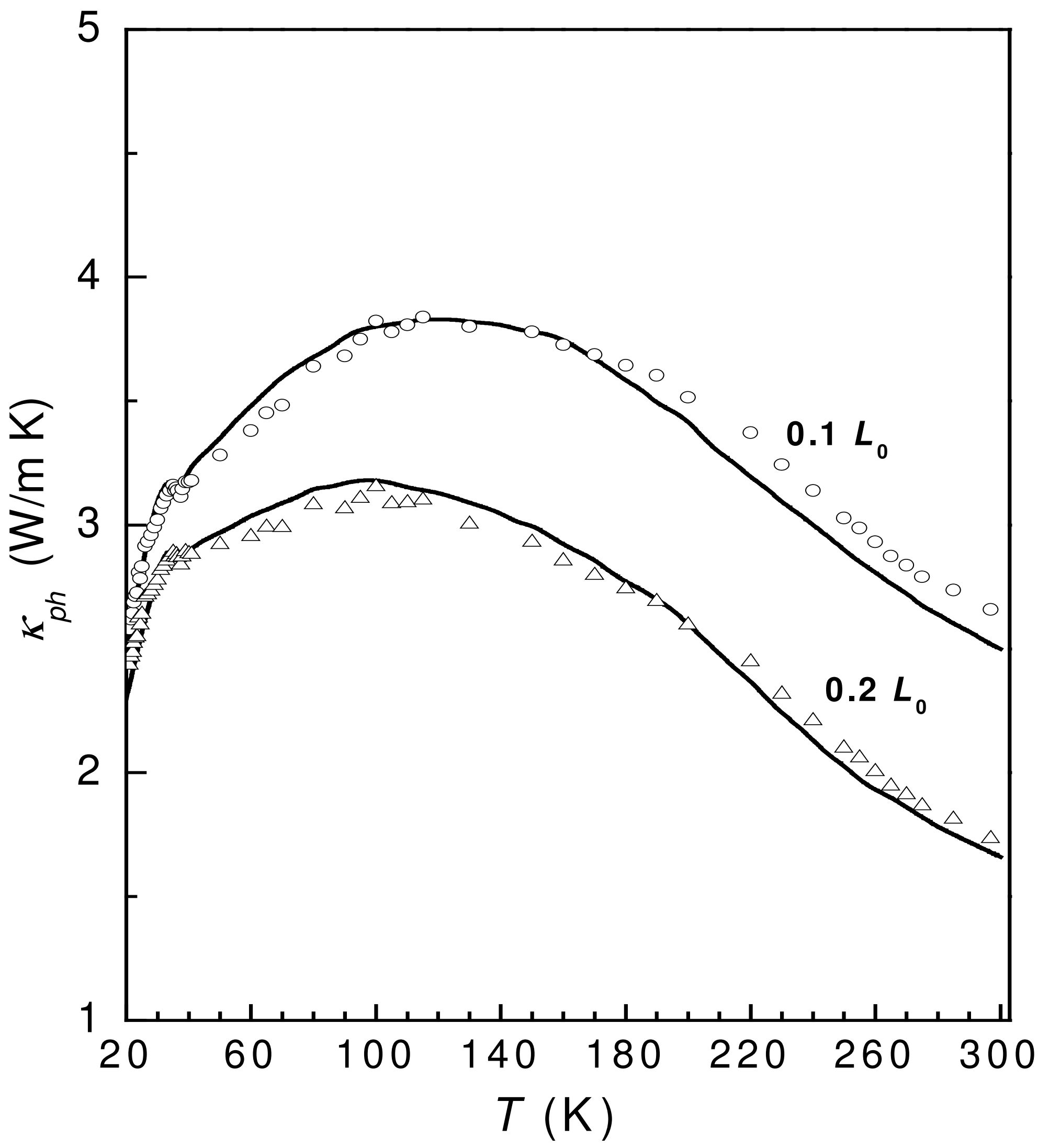